\documentclass[aps,preprint,floatfix,amsmath,superscriptaddress,endfloats,byrevtex]{revtex4}
\usepackage{graphicx}

\begin{document}

\title{Variational approach to the ground state of an impurity in Bose-Einstein condensate}

\author{Alexey Novikov}
\author{Mikhail Ovchinnikov}
\affiliation{Department of Chemistry, University of Rochester, RC
Box 270216,
   Rochester, NY 14627-0216, USA}

\date{\today}

\begin{abstract}

In this paper we consider the effect of self-localization of a
quantum impurity in Bose-Einstein condensate. Space correlation
function of the impurity is evaluated with the help of the
imaginary-time path integral approach. Employing the Feynman's
variational method we calculate the impurity correlation function as
well as the energy of the system associated with the impurity. The
effect of self-localization predicted before within Gross-Pitaevskii
approach is recovered by our treatment. The strong coupling regime
with negative ground state energy is reached by variational method,
and corresponding correlation function is calculated.

\end{abstract}
\maketitle
\section{Introduction}
\label{intro}

A theory of atomic size impurity in Bose Einstein fluids has become a subject of intensive research in the last decade.
 The theoretical development is largely motivated by recent experiments.  MIT group \cite{chikkatur2000} has studied the dynamics of impurity
 atoms in Bose Einstein condensate (BEC) of ultracold atoms in magnetic traps.  Another set of experiments studies the
microscopic superfluidity of liquid helium by spectroscopic
mesurements on molecules imbedded in helium droplets
\cite{vilesov1998, vilesov2000}. It is generally observed that
microscopic impurity particle interacting with Bose liquid/gas
behaves as a free particle with effective mass increased by its
interaction with Bose fluid. Such dissipationless quantum motion is
observed both for translational motion of particles in BEC and for
rotations of molecules in superfluid helium.

A question that received attention of theorists has been the
structure of the ground state of impurity in BEC, in particular, the
possibility of the so called self-localization of an impurity: the
appearance of a bound state of an impurity with BEC despite the
purely repulsive interaction potential.  Such behavior has been
investigated by a number of authors using linearized
Gross-Pitaevskii equations \cite{kalas2006, cucchietti2006,
sacha2006, bruderer08}. This method allows to obtain the non-linear
imaginary time Schr\"odinger equation for the impurity particle. The
energy of the particle becomes negative above certain value of
coupling strength that depends on the BEC/particle and BEC/BEC
interaction potentials. Extension of these results within fully
quantum treatment is an open question and is the main subject of
present work.

In this paper the matrix element of the reduced density operator of
impurity is calculated which allows to compute the energy and the
space correlation function of impurity.  The system of our interest
is the particle interacting with the gas of uncoupled Bogoliubov's
excitations. We formulate this matrix element via imaginary time
path integrals. The integral over BEC trajectories is explicitly
eliminated and the non-Gaussian functional integral over impurity
trajectories is left to be calculated.    We treat this integral
using the variational approach developed by Feynman for the polarons
in polar crystals \cite{feynman55}. This method was successfully
applied to the problems described by the Fr\"ohlich-type
Hamiltonians such as polarons, electron-plasmon interaction
\cite{lakhno98}, nucleon-meson interaction \cite{rosenfelder95},
many body fermion problems \cite{devreese01}, and many others. Since
the Hamiltonian of our model (one particle interacting with
uncoupled Bosonic modes) is of the Fr\"ohlich type, it lets us
believe that such treatment is successful in our case. We obtain the
free energy of an impurity as a function of coupling constant.  The
energy is increasing in the weak-coupling regime in agreement with
the regular perturbation theory treatments. With further increase of
coupling it reaches maximum and decreases becoming negative in the
case of strong-coupling. This indicates the existence of the bound
state. By computing the correlation function we obtain the
localization radius as a function of the coupling constant. We show
that the critical localization radius is of the order of magnitude
of the inverse critical momentum above which the dissipation of
particle motion takes place in real time dynamics.

The Plank constant $\hbar$ and the Boltzmann constant $k_B$ are set
to unity throughout the paper.

\section{MODEL HAMILTONIAN AND statement of a problem}


\label{model}

As it has been mentioned in the Sec.~\ref{intro}, we are going to
concentrate on the case of dilute Bose gas at the temperature mach
less than a temperature of condensation. So the gas of weakly
interacting Bose particles with mass $m$ can be described as a gas
of uncoupled Bogoluibov's excitations \cite{bogol1947, AGD}, i.e.
the Hamiltonian of BEC has diagonal form and reads
\begin{eqnarray}
H_B=\sum_{\bf k}\epsilon(k)\hat{B}^+_{\bf k}\hat{B}_{\bf k}~,~~~~~
\epsilon(k)=\sqrt{\frac{k^2}{2m_B}\left(\frac{k^2}{2m_B}+2m_Bc^2\right)}~.
\end{eqnarray}
Here bosonic operators $B_{\bf k}^+$ and $B_{\bf k}$ create and
annihilate the collective excitation with momentum ${\bf k}$ and
with the Bogoliubov's spectrum $\epsilon(k)$ which has the
phonon-like behavior at low momenta, i.e. $\epsilon(k\to0)=kc$,
where $c$ is the speed of sound.

Next, we consider a quantum particle with mass $M$ interacting with
the BEC discussed above. The Hamiltonian of the whole system has the
form \cite{miller62}
\begin{eqnarray}
\label{fullham} H&=&gn+H_B+\sum_{\bf q}\frac{{\bf
q}^2}{2M}\hat{a}_{\bf q}^+\hat{a}_{\bf q}
+\sum_{{\bf q},{\bf k}\not= 0}\gamma_{\bf
k}\left(\hat{a}^+_{{\bf q}-{\bf k}}\hat{a}_{\bf q}\hat{B}^+_{\bf
k}+\hat{a}^+_{{\bf q}+{\bf k}}\hat{a}_{\bf q}\hat{B}_{\bf
k}\right)~,
\\
\gamma_{\bf
k}&=&\frac{g}{V}\sqrt{\frac{Nk^2}{2m\epsilon(k)}}~.\nonumber
\end{eqnarray}
The bosonic operators $a^+_{\bf q}$ and $a_{\bf q}$ in the above
expression create and annihilate a single particle with the mass $M$
in the state with momentum ${\bf p}$. The last term in
Eq.~(\ref{fullham}) describes the particle-BEC interaction with the
coupling constant $\gamma_{\bf q}$ depending on whole number of
particles in bose gas $N$ in the volume $V=N/n_0$. Also, the
coupling strength depends on the coupling constant, $g$, defined as
a zero Fourier component of inpurity/BEC interaction potential. This
constant is related to the speed of sound as $c=\sqrt{gn_0/m_B}$.
The first term in (\ref{fullham}) represents the first order
correction to the energy due to the particle-BEC interaction. In the
interaction part of the Hamiltonian as well as in the Hamiltonian of
the free BEC we neglected the terms responsible for the interaction
between the Bogoliubov's excitations. This approximation remains
valid if the single impurity alters the surrounding BEC only
slightly which is always the case in the macroscopic limit.

In order to describe the statistical properties of the impurity that
stays in thermal equilibrium with the BEC environment, we will
calculate the space correlation function defined as a matrix element
of reduced density operator
\begin{eqnarray}
C({\bf x},{\bf x}^\prime)=\frac{1}{Z}\langle{\bf x}|{\rm
Tr}_Be^{-\beta H}|{\bf x}^\prime\rangle~.
\end{eqnarray}
Here $\beta$ is the inverse temperature, the Hamiltonian $H$ is
defined by Eq.~(\ref{fullham}), the trace is performed over all
states of BEC and Z is the partition function. Due to the momentum
conservation in the whole system the reduced density operator is
diagonal in momentum space. Thus
\begin{eqnarray}
C({\bf x}-{\bf x}^\prime)=\frac{1}{ZV}\sum_{\bf p}e^{-i{\bf p}({\bf
x}-{\bf x^\prime})}\langle{\bf p}|{\rm Tr}_Be^{-\beta H}|{\bf
p}\rangle=\frac{1}{V}\sum_{\bf p}e^{-i{\bf p}({\bf x}-{\bf
x^\prime})}C_{\bf p}~.
\end{eqnarray}
So our task is to calculate the correlation function as the reduced
density matrix element between two states of the impurity with the
momentum ${\bf p}$
\begin{eqnarray}
C_{\bf p}=\rho_{\bf p}(\beta)=\frac{1}{Z}{\rm Tr}_B\langle{\bf
p}|e^{-\beta H}|{\bf p}\rangle=\frac{1}{Z}{\rm Tr}_B\langle
0|\hat{a}_{\bf p}e^{-\beta H}\hat{a}^+_{\bf p}|0\rangle
\end{eqnarray}
The above matrix element can be represented as the imaginary-time
coherent-state functional integral \cite{novikov2009spectr}
\begin{eqnarray}
\label{f1}
C_{\bf p}(\beta)&=&\frac{1}{Z}\int D[\{a_{\bf
q}^*(\tau)\},\{a_{\bf q}(\tau)\}]\int\prod_{\bf k}\frac{db_{\bf
k}^*db_{\bf k}}{\pi}\int D[\{b_{\bf k}^*(\tau)\},\{b_{\bf
k}(\tau)\}]a_{\bf p}(\beta)a_{\bf p}^*(0)\nonumber
\\
&\times&\exp\left[-\sum_{\bf k}|b_{\bf k}|^2+\sum_{\bf k}b_{\bf
k}^*(\beta)b_{\bf k}(\beta)-S\right]
\end{eqnarray}
with the imaginary-time action of the whole system
\begin{eqnarray}
S&=&\int_0^\beta d\tau\left[\sum_{\bf q}\big(\dot{a}_{\bf
q}(\tau)a^*(\tau)+E({\bf q})a_{\bf q}^*(\tau)a_{\bf
q}(\tau)\big)+\sum_{\bf k}\big(\dot{b}_{\bf
k}(\tau)b^*(\tau)+\epsilon({\bf k})b_{\bf k}^*(\tau)b_{\bf
k}(\tau)\big)\right.\nonumber
\\
&+&\left.\sum_{\bf q,k}\gamma_{\bf k}\big(a^*_{\bf q+k}(\tau)a_{\bf
q}(\tau)b_{\bf k}(\tau)+a^*_{\bf q-k}(\tau)a_{\bf q}(\tau)b^*_{\bf
k}(\tau)\big)\right]~.
\end{eqnarray}
The integral (\ref{f1}) must be evaluated with the following
boundary conditions for the trajectories
\begin{eqnarray}
b_{\bf k}(0)&=&b_{\bf k}~,~~~b_{\bf k}^*(\beta)=b^*_{\bf k}~,\\
a_{\bf q}(0)&=&a^*_{\bf q}(\beta)=0~.
\end{eqnarray}
Because of the linear dependence of the interaction part of the
action on the $b_{\bf k}(\tau),~b^*_{\bf k}(\tau)$ trajectories the
BEC degrees of freedom can be integrated out immediately using the
formula
\begin{eqnarray}
\label{elimination}
&&\oint D[\{b_{\bf k}^*(\tau)\},\{b_{\bf
k}(\tau)\}]\exp\left[-\sum_{\bf k}\int_0^\beta
d\tau\Big(\epsilon({\bf k})b_{\bf k}^*(\tau)b_{\bf k}(\tau)-j_{\bf
k}(\tau)b_{\bf k}(\tau)-j^*_{\bf k}(\tau)b^*_{\bf
k}(\tau)\Big)\right] \nonumber
\\
&&=Z_B\exp\left(\sum_{\bf k}\int_0^\beta d\tau\int_0^\beta
d\tau^\prime\Gamma_{\bf k}(\tau-\tau^\prime)j_{\bf k}(\tau)j^*_{\bf
k}(\tau^\prime)\right)~.
\end{eqnarray}
The function $\Gamma_{\bf k}(\tau)$ represents the imaginary-time
propagator of free Bose field with the spectrum $\epsilon({\bf k})$
\begin{eqnarray}
\label{propagatorgamma}
\Gamma_{\bf
k}(\tau)=\big(\Theta(\tau)+n_{\bf k}\big)e^{-\tau\epsilon({\bf
k})}~,~~~~n_{\bf k}=\frac{1}{e^{\beta\epsilon({\bf k})}-1}~,
\end{eqnarray}
where function $\Theta(\tau)$ denotes the Heaviside step function,
and $Z_B=\prod_{\bf k}n_{\bf k}e^{\beta\epsilon({\bf k})}$ is its
partition function.

Thus, after tracing out the BEC we obtain the functional integral
representation of the reduced density matrix of the relevant
particle that includes the integration over the particle
trajectories only
\begin{eqnarray}
\label{f2}
\rho_{\bf p}(\beta)=\frac{Z_B}{Z}\int D[\{a_{\bf
q}^*(\tau)\},\{a_{\bf q}(\tau)\}]a_{\bf p}(\beta)a_{\bf
p}^*(0)\exp\left[-S_P+S_I\right]
\end{eqnarray}
Here we defined the free imaginary-time action of relevant particle
\begin{eqnarray}
S_P=\int_0^\beta d\tau\sum_{\bf q}\big(\dot{a}_{\bf q}(\tau)a^*_{\bf
q}(\tau)+E({\bf q})a_{\bf q}^*(\tau)a_{\bf q}(\tau)\big)
\end{eqnarray}
The integral (\ref{f2}) is of the non-Gaussian type and its
non-Gaussian part $S_I$ is due to the impurity-BEC coupling and
reads
\begin{eqnarray}
\label{si}
S_I=\int_0^\beta d\tau\int_0^\beta d\tau^\prime\sum_{\bf
q,q^\prime,k}\gamma_{\bf k}^2a^*_{\bf q+k}(\tau)a_{\bf
q}(\tau)\Gamma_{\bf k}(\tau-\tau^\prime)a^*_{\bf
q^\prime-k}(\tau^\prime)a_{\bf q^\prime}(\tau^\prime)~.
\end{eqnarray}

Having the action of the impurity, we can investigate its classical
dynamics described by corresponding Euler's equations of motion. For
this purpose let us write the action in the coordinate
representation by changing the variables of the functional
integration as $a_{\bf k}(\tau)=\frac{1}{\sqrt{V}}\int d{\bf
r}e^{-i{\bf kr}}\psi({\bf r},\tau)$. Now the action of the impurity
surrounded by BEC has the form
\begin{eqnarray}
S_{imp}&=&S_P-S_I=\int_0^\beta\int d{\bf r}\psi^*({\bf
r},\tau)\left(\frac{\partial}{\partial\tau}-\frac{\Delta}{2M}\right)\psi({\bf
r},\tau)-
\\
&&\int_0^\beta d\tau\int_0^\beta d\tau^\prime\int d{\bf r}\int d{\bf
r}^\prime\psi^*({\bf r},\tau)\psi({\bf r},\tau)\Lambda({\bf
r-r^\prime},\tau-\tau^\prime)\psi^*({\bf
r^\prime},\tau^\prime)\psi({\bf r^\prime},\tau^\prime)\nonumber~,
\end{eqnarray}
\begin{eqnarray}
\label{lambda}
\Lambda({\bf
r-r^\prime},\tau-\tau^\prime)=\frac{V}{(2\pi)^3}\int d{\bf
k}\gamma_{\bf k}^2\Gamma_{\bf k}(\tau-\tau^\prime)e^{i{\bf kr}}
\end{eqnarray}
Varying the above action one gets the following equations of motion
\begin{eqnarray}
\dot{\psi}({\bf r},\tau)&=&\frac{\Delta}{2M}\psi({\bf r},\tau)+\int
d\tau^\prime\int d{\bf r^\prime}\psi({\bf
r},\tau)\tilde{\Lambda}({\bf
r-r^\prime},\tau-\tau^\prime)\psi^*({\bf
r^\prime},\tau^\prime)\psi({\bf r^\prime},\tau^\prime)~,
\\
-\dot{\psi}^*({\bf r},\tau)&=&\frac{\Delta}{2M}\psi^*({\bf
r},\tau)+\int d\tau^\prime\int d{\bf r^\prime}\psi^*({\bf
r},\tau)\tilde{\Lambda}({\bf
r-r^\prime},\tau-\tau^\prime)\psi^*({\bf
r^\prime},\tau^\prime)\psi({\bf r^\prime},\tau^\prime)~,\nonumber
\end{eqnarray}
with symmetrized kernel
\begin{eqnarray}
&&\tilde{\Lambda}({\bf r-r^\prime},\tau-\tau^\prime)=\Lambda({\bf
r-r^\prime},\tau-\tau^\prime)+\Lambda({\bf
r^\prime-r},\tau^\prime-\tau)~.\nonumber
\end{eqnarray}
We will seek the stationary solutions in the form $\psi({\bf
r},\tau)=\psi_s({\bf r})e^{-\tau E}$ and $\psi^*({\bf
r},\tau)=\psi_s^*({\bf r})e^{\tau E}$. So the stationary equation
for the impurity wave function is
\begin{eqnarray}
\label{stateqs} E\psi_s({\bf r})+\frac{\Delta}{2M}\psi_s({\bf
r})+\int d{\bf r^\prime}\psi_s({\bf r})\psi^*_s({\bf
r^\prime})\psi_s({\bf r^\prime})\Lambda_s({\bf r-r^\prime})=0~,
\end{eqnarray}
where
\begin{eqnarray}
\Lambda_s({\bf r-r^\prime})=\int d\tau^\prime\tilde{\Lambda}({\bf
r-r^\prime},\tau-\tau^\prime)~.
\end{eqnarray}

Using the definition of the BEC-propagator
Eq.~(\ref{propagatorgamma}) together with the Eq.~(\ref{lambda}) one
obtains the non-linear kernel $\Lambda_s$, given by
\begin{eqnarray}
\label{lambdas}
\Lambda_s({\bf
r})=\frac{m_Bn_0g^2}{\pi}\frac{e^{-2m_Bcr}}{r}~.
\end{eqnarray}
The equations (\ref{stateqs}) with the kernel (\ref{lambdas})
represent the particle self-interacting through the screened Coulomb
potential. The same equations for the impurity in weak-coupled BEC
can also be obtained directly with the help of linearized
Gross-Pitaevskii equations with more restricting initial assumptions
\cite{cucchietti2006} such as Hartree approximation for the
impurity/BEC wave function and using the real-valued BEC wave
function. In Ref.~\cite{cucchietti2006} authors investigated the
lowest energy solution of the above equations. They found that the
equations (\ref{stateqs}) have the localized solution with negative
energy in the strong coupling regime, i.e. when the impurity/BEC
interaction is strong enough to compensate the high kinetic energy
of the localized state of the impurity.

The aim of this paper is to construct fully quantum description of
the model system based on the calculation of the imaginary-time
quantum propagator. The quantum formulation of the problem without
the restriction of a mean-field approximation will be given in the
following section.

\section{Path integral formulation of the correlation function}

Our goal is to describe the effect of self-localization from purely
quantum point of view. For this purpose let us return to the
functional integral for the correlation function with the full
impurity-plus-BEC action of the form (\ref{f1}). Our aim is to
represent the correlation function as a single Feynman's path
integral over the impurity trajectories. As in the previous section
we will use the coordinate representation for impurity trajectories.
So the correlation function is given by
\begin{eqnarray}
C({\bf x-x^\prime},\beta)=\frac{1}{Z}\oint D[\{b^*_{\bf
k}(\tau)\},\{b_{\bf k}(\tau)\}]\int D[\{\psi^*({\bf
r},\tau)\},\{\psi({\bf r},\tau)\}]\psi({\bf
x^\prime},\beta)\psi^*({\bf x},0)e^{-S}~,
\end{eqnarray}
where the action is
\begin{eqnarray}
S&=&\int_0^\beta d\tau\sum_{\bf k}\big(\dot{b}_{\bf
k}(\tau)b^*(\tau)+\epsilon({\bf k})b_{\bf k}^*(\tau)b_{\bf
k}(\tau)\big)+\int_0^\beta d\tau\int d{\bf r}\left(\psi^*({\bf
r},\tau)\dot{\psi}({\bf r},\tau)-\psi^*({\bf
r},\tau)\frac{\Delta}{2M}\psi({\bf r},\tau)\right)\nonumber
\\
&\times&\int_0^\beta d\tau\int d{\bf r}\psi^*({\bf r},\tau)\psi({\bf
r},\tau)\sum_{\bf k}\gamma_{\bf k}\left(b_{\bf k}(\tau)e^{i{\bf
kr}}+b^*_{\bf k}(\tau)e^{-i{\bf kr}}\right)~.
\end{eqnarray}
Let us introduce auxiliary external sources into the action and
define the following functional
\begin{eqnarray}
\label{genfuncf}
&&F[j^*({\bf r},\tau),j({\bf r},\tau)]=\int
D[\{\psi^*({\bf r},\tau)\},\{\psi({\bf r},\tau)\}]
\\
&&\times\exp\left[-\int_0^\beta d\tau\int d{\bf r}\left(\psi^*({\bf
r},\tau)\hat{K}({\bf r},\tau)\psi({\bf r},\tau)-j({\bf
r},\tau)\psi({\bf r},\tau)-j^*({\bf r},\tau)\psi^*({\bf
r},\tau)\right)\right]~,\nonumber
\end{eqnarray}

\begin{equation}
\hat{K}({\bf
r},\tau)=\frac{\partial}{\partial\tau}-\frac{\Delta}{2M}+\sum_{\bf
k}\gamma_{\bf k}\left(b_{\bf k}(\tau)e^{i{\bf kr}}+b^*_{\bf
k}(\tau)e^{-i{\bf
kr}}\right)=\frac{\partial}{\partial\tau}+{\hat{\mathcal H}}({\bf
r},\tau)~.
\end{equation}
The functional integral in the right side of Eq.~(\ref{genfuncf})
has to be evaluated with the boundary conditions $\psi({\bf
r},0)=\psi^*({\bf r},\beta)=0$. Since the integral in
(\ref{genfuncf}) is Gaussian, it can be done by the stationary phase
method. Thus for the stationary trajectories one finds
\begin{eqnarray}
\dot{\psi}({\bf r},\tau)-{\hat{\mathcal H}}({\bf r},\tau)\psi({\bf
r},\tau)-j^*({\bf r},\tau)=0~,
\\
-\dot{\psi}^*({\bf r},\tau)-{\hat{\mathcal H}}({\bf
r},\tau)\psi^*({\bf r},\tau)-j({\bf r},\tau)=0~.\nonumber
\end{eqnarray}
The formal solution of the above equations can be written in the
following form
\begin{eqnarray}
\psi_s({\bf r},\tau)=\int_0^\tau
d\tau^\prime\exp\left(\int_{\tau^\prime}^{\tau}{\hat{\mathcal
H}}({\bf r},s)ds\right)j^*({\bf r},\tau^\prime)~.
\end{eqnarray}
Finally, substituting this solution into the integrand in
(\ref{genfuncf}), for the functional $F$ we have
\begin{eqnarray}
F[j^*({\bf r},\tau),j({\bf r},\tau)]&=&\exp\left(\int_0^\beta
d\tau\int d{\bf r}j({\bf r},\tau)\psi_s({\bf r},\tau)\right)
\\
&=&\exp\left(\int_0^\beta d\tau\int_0^\beta
d\tau^\prime\Theta(\tau-\tau^\prime)\langle j({\bf
r},\tau)|e^{-\int_{\tau^\prime}^\tau\hat{{\mathcal H}}({\bf
r},s)ds}|j^*({\bf r},\tau^\prime)\rangle\right)\nonumber
\end{eqnarray}
Now we can note that the correlation function can be written with
the help of the generating functional, namely
\begin{eqnarray}
\label{c10}
C({\bf
x-x^\prime},\beta)=\left\langle\frac{\delta^2}{\delta j({\bf
x^\prime},\beta)\delta j^*({\bf x},0)}F[j^*({\bf r},\tau),j({\bf
r},\tau)]\right\rangle\Bigg|_{j=j^*=0}~,
\end{eqnarray}
where the averaging is performed as the integration over BEC
trajectories with the free action of BEC as
\begin{eqnarray}
\langle ...\rangle=\frac{1}{Z}\oint D[\{b^*_{\bf k}(\tau)\},\{b_{\bf
k}(\tau)\}]~...~\exp\left[\int_0^\beta d\tau\sum_{\bf
k}\big(\dot{b}_{\bf k}(\tau)b^*(\tau)+\epsilon({\bf k})b_{\bf
k}^*(\tau)b_{\bf k}(\tau)\big)\right]~.
\end{eqnarray}
Calculating the functional derivative in Eq.~(\ref{c10}) and writing
one particle propagator $\langle j({\bf
r},\tau)|e^{-\int_{\tau^\prime}^\tau\hat{{\mathcal H}}({\bf
r},s)ds}|j^*({\bf r},\tau^\prime)\rangle$ as Feynman's path
integral, for the correlation function we have
\begin{eqnarray}
\label{int30}
C({\bf x-x^\prime},\beta)=\left\langle\int_{{\bf
x}(0)={\bf x}}^{{\bf x}(\beta)={\bf x^\prime}}D[{\bf
x}(\tau)]e^{-S_p} \right\rangle~,
\end{eqnarray}
where the single particle action depends on the coordinate
trajectory ${\bf x}(\tau)$
\begin{eqnarray}
S_p=\int_0^\beta d\tau\left[\frac{M\dot{\bf x}^2}{2}+\sum_{\bf
k}\gamma_{\bf k}\left(b_{\bf k}(\tau)e^{i{\bf k}{\bf
x}(\tau)}+b^*_{\bf k}(\tau)e^{-i{\bf k}{\bf x}(\tau)}\right)\right]
\end{eqnarray}
Now using the formula (\ref{elimination}) we can eliminate the
integration over BEC trajectories from the equation (\ref{int30})
and get the correlation function as a single Feynman's path integral
for the impurity
\begin{eqnarray}
\label{int31}
C({\bf x-x^\prime},\beta)=\int_{{\bf x}(0)={\bf
x}}^{{\bf x}(\beta)={\bf x^\prime}}D[{\bf x}(\tau)]e^{-S_R}~,
\end{eqnarray}
where the impurity action reads
\begin{eqnarray}
\label{realact} S_R=\int_0^\beta d\tau\frac{M\dot{\bf
x}^2}{2}-\int_0^\beta d\tau\int_0^\beta d\tau^\prime\sum_{\bf
k}\gamma_{\bf k}^2\Gamma_{\bf k}(\tau-\tau^\prime)e^{i{\bf k}({\bf
x}(\tau)-{\bf x}(\tau^\prime))}~.
\end{eqnarray}
This form of the single particle functional integral is considered
in the following section.

\section{Feynman's variational approach}

The imaginary-time correlation function in the form of path integral
(\ref{int31}) provides the exact description of the statistics of
the impurity surrounded by the degenerate BEC at low temperature. In
this section we will calculate this integral approximately using
Feynman's variational approach to the polaron problem
\cite{feynman55}. The idea of the original method is to replace the
real action of the impurity (\ref{realact}) with the trial action of
the form
\begin{eqnarray}
\label{trialact}
S_T=\int_0^\beta d\tau\frac{M\dot{\bf
x}^2}{2}+\int_0^\beta d\tau\int_0^\beta d\tau^\prime
Q(\tau-\tau^\prime)\big({\bf x}(\tau)-{\bf x}(\tau^\prime)\big)^2~,
\end{eqnarray}
where
\begin{eqnarray}
Q(\tau-\tau^\prime)=Q_0\big(\Theta(\tau-\tau^\prime)+n_\omega\big)e^{-\omega(\tau-\tau^\prime)}~,~~~~n_\omega=\frac{1}{e^{\omega\beta}-1}~.
\end{eqnarray}
It is also useful to write  the trial action in the extended form as
\begin{eqnarray}
\label{actext}
e^{-S_T}&\sim&\oint D[{\bf y}(\tau)]e^{-S_{ET}}~,
\\
S_{ET}&=&\int_0^\beta d\tau\left[\frac{M\dot{\bf
x}^2(\tau)}{2}+\frac{m\dot{\bf
y}^2(\tau)}{2}+\frac{m\omega^2\big({\bf y}(\tau)-{\bf
x}(\tau)\big)^2}{2}\right]~.
\end{eqnarray}
Performing the integration over trajectories ${\bf y}(\tau)$ one
recovers the trial action in the form of Eq.~(\ref{trialact}) with
$Q_0=m\omega^3/4$. Thus in this variational treatment we replace the
interaction with the original BEC environment by the interaction
with single trial particle with the mass $m$.

The trial action of the form (\ref{trialact}) is assumed to be the
zeroth order contribution while the difference between real action
(\ref{realact}) and trial one has to be considered as a
perturbation. Thus one has to write the following expansion of the
correlation function
\begin{eqnarray}
\label{ct}
C_T({\bf x-x^\prime},\beta)=\frac{\langle
1+S_T-S_R\rangle}{\langle 1+S_T-S_R\rangle_0}.
\end{eqnarray}
where we have defined two kinds of averages
\begin{eqnarray}
\langle ...\rangle=\int_{{\bf x}(0)={\bf x}}^{{\bf x}(\beta)={\bf
x^\prime}}D[{\bf x}(\tau)]...~e^{-S_T},~~~~\langle
...\rangle_0=\int_{{\bf x}(0)=0}^{{\bf x}(\beta)=0}D[{\bf
x}(\tau)]...~e^{-S_T}~.
\end{eqnarray}
The trial correlation function $C_T$ in Eq.~(\ref{ct}) still depends
on two variational parameters $Q_0$ and $\omega$ (or $m$ and
$\omega$). Thus, in accordance with the principle of minimal
sensitivity \cite{kleinertbook} one has to minimize this expansion
with respect to the variational parameters, and the extremum point
will give the best variational approximation for the path integral
(\ref{int31}). In order to proceed with the expansion (\ref{ct}) we
will have to calculate four functions, namely
\begin{eqnarray}
\label{sigma} \Sigma_1&=&\int_0^\beta d\tau\int_0^\beta d\tau^\prime
Q(\tau-\tau^\prime)\big\langle\big({\bf x}(\tau)-{\bf
x}(\tau^\prime)\big)^2\big\rangle~,\nonumber
\\
\Sigma_2&=&\int_0^\beta d\tau\int_0^\beta d\tau^\prime\sum_{\bf
k}\Gamma_{\bf k}(\tau-\tau^\prime)\big\langle\exp\left[i{\bf
k}\big({\bf x}(\tau)-{\bf x}(\tau^\prime)\big)\right]\big\rangle~,
\end{eqnarray}
and
\begin{eqnarray}
\label{sigma0} \Sigma_1^0&=&\int_0^\beta d\tau\int_0^\beta
d\tau^\prime Q(\tau-\tau^\prime)\big\langle\big({\bf x}(\tau)-{\bf
x}(\tau^\prime)\big)^2\big\rangle_0~,\nonumber
\\
\Sigma_2^0&=&\int_0^\beta d\tau\int_0^\beta d\tau^\prime\sum_{\bf
k}\Gamma_{\bf k}(\tau-\tau^\prime)\big\langle\exp\left[i{\bf
k}\big({\bf x}(\tau)-{\bf x}(\tau^\prime)\big)\right]\big\rangle_0~,
\end{eqnarray}
The expansion (\ref{ct}) in terms of functions $\Sigma$ becomes
\begin{eqnarray}
C_T=\frac{\langle 1\rangle+\Sigma_1+\Sigma_2}{\langle
1\rangle_0+\Sigma_1^0+\Sigma_2^0}
\end{eqnarray}

In order to calculate the average $\big\langle\big({\bf
x}(\tau)-{\bf x}(\tau^\prime)\big)^2\big\rangle$ and
$\big\langle\exp\left[i{\bf k}\big({\bf x}(\tau)-{\bf
x}(\tau^\prime)\big)\right]\big\rangle$ that enters the functions
$\Sigma_1$ and $\Sigma_2$, respectively, one can use the generating
functional of the form
\begin{eqnarray}
\label{genw1}
W[{\bf j} (\tau)]=\oint D[{\bf y}(\tau)]\int_{{\bf
x}(0)={\bf x}}^{{\bf x}(\beta)={\bf x^\prime}}D[{\bf
x}(\tau)]\exp\left(-S_{ET}+\int_0^\beta ds{\bf x}(s){\bf
j}(s)\right)
\end{eqnarray}
So the second average $\langle\exp\left[i{\bf k}\big({\bf
x}(\tau)-{\bf x}(\tau^\prime)\big)\right]\rangle$ can be evaluated
by setting ${\bf j(s)}=i{\bf
k}\big(\delta(s-\tau)-\delta(s-\tau^\prime)\big)$ while the first
one is obtained by differentiating the second one twice over ${\bf
k}$ at ${\bf k}=0$. The functional integral in (\ref{genw1}) can be
evaluated by introducing the new variables ${\bf q}(\tau)={\bf
x}(\tau)-{\bf y}(\tau)$ and ${\bf r}(\tau)=\big(M{\bf x}(\tau)+m{\bf
y}(\tau)\big)/(M+m)$. Substituting this replacement into
(\ref{actext}) one gets the action $S_{ET}$ in the form
\begin{eqnarray}
S_{ET}&=&\int_0^\beta d\tau\left[\frac{(M+m)\dot{\bf
r}^2(\tau)}{2}+\frac{\mu\dot{\bf q}^2(\tau)}{2}+\frac{m\omega^2{\bf
q}^2(\tau)}{2}-{\bf j}(\tau){\bf r}(\tau)-\frac{\mu}{M}{\bf
j}(\tau){\bf q}(\tau)\right]~,
\\
\mu&=&\frac{mM}{m+M}~. \nonumber
\end{eqnarray}
The above action now describes harmonic oscillator and free particle
in presence of the external source and the corresponding functional
integral can be easily evaluated. Here we will write down the result
of the integration in (\ref{genw1})
\begin{eqnarray}
\label{generating}
&&W[{\bf j} (\tau)]({\bf x},{\bf
x^\prime})=W_0\exp\left\{-({\bf x}-{\bf
x^\prime})^2\left[\frac{M\mu}{2\beta
m}+\frac{\Omega\mu}{4}\coth\frac{\Omega\beta}{2}\right]\right.
\\
\nonumber \\
 &&+\int_0^\beta d\tau{\bf j}(\tau)\left[({\bf
x^\prime}-{\bf x})\frac{\mu}{m}\frac{\tau}{\beta}+\frac{\mu}{m}{\bf
x}+\frac{\mu}{2M}({\bf x}+{\bf
x^\prime})+\frac{\mu}{2M}\frac{\sinh\Omega\tau-\sinh\Omega(\beta-\tau)}{\sinh\Omega\beta}({\bf
x^\prime}-{\bf x})\right]\nonumber
\\
\nonumber
\\
&&+\int_0^\beta d\tau\int_0^\beta d\tau^\prime{\bf j}(\tau){\bf
j}(\tau^\prime)\left[\frac{\tau^\prime}{2\beta}\frac{\beta-\tau}{M+m}\theta(\tau-\tau^\prime)+
\frac{\tau}{2\beta}\frac{\beta-\tau^\prime}{M+m}\theta(\tau^\prime-\tau)\right.\nonumber
\\
\nonumber
\\
&&+\frac{\mu}{4\Omega
M^2}\frac{1}{\tanh\frac{\Omega\beta}{2}}\left(\frac{\sinh\Omega(\beta-\tau)+\sinh\Omega\tau+
\sinh\Omega(\beta-\tau^\prime)+\sinh\Omega\tau^\prime}{\sinh\Omega\beta}-1\right)\nonumber
\\
\nonumber
\\
&&\left.\left.+\frac{\mu}{4\Omega
M^2}\frac{\cosh\Omega\left(\beta/2-|\tau-\tau^\prime|\right)}{\sinh\frac{\Omega\beta}{2}}\right]\right\}~,
\end{eqnarray}
where we have introduced the frequency $\Omega=\omega\sqrt{m/\mu}$.
The prefactor $W_0$ is the constant coming from the integration over
the deviations around the stationary trajectories, so it is
independent on the end-points ${\bf x}$ and ${\bf x^\prime}$.

So for the second average $\langle\exp\left[i{\bf k}\big({\bf
x}(\tau)-{\bf x}(\tau^\prime)\big)\right]\rangle$ one gets
\begin{eqnarray}
\label{exponenta}
&&\langle\exp\left[i{\bf k}\big({\bf x}(\tau)-{\bf
x}(\tau^\prime)\big)\right]\rangle=W_0\exp\left\{-({\bf x}-{\bf
x^\prime})^2\left[\frac{M}{2\beta}\frac{\omega^2}{\Omega^2}
+\frac{\Omega\mu}{4}\coth\frac{\Omega\beta}{2}\right]\right.
\\
\nonumber
\\
&&+i{\bf k}({\bf x^\prime}-{\bf
x})\left[\frac{\omega^2}{\Omega^2}\frac{\tau-\tau^\prime}{\beta}
+\frac{1}{2}\frac{\Omega^2-\omega^2}{\Omega^2}\frac{\sinh\Omega\tau-\sinh\Omega(\beta-\tau)
-\sinh\Omega\tau^\prime+\sinh\Omega(\beta-\tau^\prime)}{\sinh\Omega\beta}\right]\nonumber
\\
\nonumber
\\
&&\left.-{\bf
k}^2\left[\frac{1}{2M}\frac{\omega^2}{\Omega^2}|\tau-\tau^\prime|\left(1-\frac{|\tau-\tau^\prime|}{\beta}\right)
+\frac{1}{2M\Omega}\frac{\Omega^2-\omega^2}{\Omega^2}\frac{\cosh\frac{\Omega\beta}{2}-
\cosh\Omega\big(\beta/2-|\tau-\tau^\prime|\big)}{\sinh\frac{\Omega\beta}{2}}\right]\right\}\nonumber
\end{eqnarray}
Below we will be always interested in low temperature limit
$\beta\to\infty$. Calculating the first average
$\big\langle\big({\bf x}(\tau)-{\bf
x}(\tau^\prime)\big)^2\big\rangle$ with the help of the above
expression, for the function $\Sigma_1$ one gets
\begin{eqnarray}
\label{sigma1}
\Sigma_1\Big|_{\beta\to\infty}=W_0\left[\frac{\Omega^2+\omega^2}{2\omega\Omega^3}({\bf
x-x^\prime})^2+\frac{3\beta}{M\omega\Omega}\right]\exp\left[-\frac{\mu\Omega}{4}({\bf
x-x^\prime})^2\right]
\end{eqnarray}
The calculation of $\Sigma_2$ can not be done in closed form and
requires some numerical calculations that will be performed in the
next sections.

\section{Energy of the impurity}

In this section we will investigate the ground state energy of the
impurity as a function of the impurity-BEC coupling strength. Let us
define the free energy of the system $E$ as
\begin{eqnarray}
\label{freeenergy}
e^{-\beta E}=\int_{{\bf x}(0)=0}^{{\bf
x}(\beta)=0}D[{\bf x}(\tau)]~e^{-S_R}=\langle e^{S_T-S_R}\rangle_0~.
\end{eqnarray}
In the zero temperature limit free energy coincides with the
ordinary energy and shows the difference between coupled and
uncoupled impurity/BEC system.

The expression (\ref{freeenergy}) up to first cumulant reads
\begin{eqnarray}
e^{-\beta E(\Omega,\omega)}= \langle 1\rangle_0\exp\frac{\langle
S_T-S_R\rangle_0}{\langle 1\rangle_0}
~,~~~~E(\Omega,\omega)=E_0+E_1+E_2~.
\end{eqnarray}
The zero energy contribution $E_0$ is
\begin{eqnarray}
\label{energy0}
e^{-\beta E_0}=\int_{{\bf x}(0)=0}^{{\bf
x}(\beta)=0}D[{\bf x}(\tau)]e^{-S_T}=W[{\bf j}=0, {\bf
x-x^\prime}=0]=W_0~,
\end{eqnarray}
and the first order contribution terms are defined as
\begin{eqnarray}
E_1=-\frac{1}{\beta W_0}\Sigma_1^0~,~~~ E_2=-\frac{1}{\beta
W_0}\Sigma_2^0~,
\end{eqnarray}
where $\Sigma_{1,2}^0$ is defined by Eq.~(\ref{sigma0}). While the
terms $E_1$ and $E_2$ can be evaluated directly using the generating
functional, the first term $E_0$ can not be calculated explicitly
from the formula (\ref{energy0}). In order to find $E_0$ one has to
note that
\begin{eqnarray}
\int_0^\beta d\tau\int_0^\beta d\tau^\prime
Q(\tau-\tau^\prime)\big\langle\big({\bf x}(\tau)-{\bf
x}(\tau^\prime)\big)^2\big\rangle_0=-m\frac{d}{dm}e^{-\beta E_0}
\end{eqnarray}
Using the result (\ref{sigma1}) we can get following equation for
zero order energy
\begin{eqnarray}
\frac{dE_0}{dm}=\frac{3}{4}\frac{\omega^2}{M\Omega}~,~~~E_0(m=0)=0.
\end{eqnarray}
The solution of the above equation is
\begin{eqnarray}
E_0=\frac{3}{2}(\Omega-\omega)~,
~~~E_0+E_1=\frac{3}{4}\frac{(\Omega-\omega)^2}{\Omega}~.
\end{eqnarray}
Now it is left to calculate the contribution $E_2$. With the help of
the formula (\ref{exponenta}), in the zero temperature limit one can
give the following expression for $E_2$
\begin{eqnarray}
E_2=-\int_0^\infty d\tau\sum_{\bf k}\gamma_{\bf
k}^2\exp\left[-\epsilon({\bf k})\tau-\frac{{\bf
k}^2}{2}\left(\frac{\omega^2}{\Omega^2}\frac{\tau}{M}+\frac{\Omega^2-\omega^2}{\Omega^2}\frac{1-e^{-\Omega\tau}}{M\Omega}\right)\right]
\end{eqnarray}

First, let us consider the weak coupling limit, i.e. $Q_0\to 0$ or
$m\to 0$ so that $\Omega\sim\omega$. Thus we have in the lowest
order $E_0+E_1=0$ and the energy $E$ is given by the first expansion
term of the ordinary perturbation theory
\begin{eqnarray}
E^{weak}=E_2^{weak}=-\sum_{\bf k}\gamma_{\bf
k}\frac{1}{\epsilon({\bf k})+\frac{k^2}{2M}}~.
\end{eqnarray}
In order to prevent the ultraviolet divergence in the above sum over
momenta one has to renormalize the coupling constant $g$ according
to the second order Born approximation. For this purpose we must add
the zero order energy term $gn$ from the Hamiltonian (\ref{fullham})
to the energy expression and expand the constant $g$ in powers of
scattering length $a$ for the impurity-BEC interaction
\begin{eqnarray}
g=\frac{2\pi a}{m_r}\left(1+\frac{2a}{\pi}\int
dk\right)~,~~~m_r=\frac{m_BM}{m_B+M}~,
\end{eqnarray}
The weak coupling energy reexpanded in powers of scattering length
in now finite and reads
\begin{eqnarray}
E^{weak}=m_Bc^2\alpha(1+z)\int_0^\infty
dl\left[1-\frac{1+z}{\sqrt{1+4/l^2}\left(\sqrt{1+4/l^2}+z\right)}\right]
\end{eqnarray}
Here we have introduced dimensionless coupling parameter
$\alpha=4an/(m_Bc)$ and the mass ratio $z=m_B/M$. Below we will
always use the renormalized expression for $E_2$ which is
\begin{eqnarray}
E_2^{ren}=E_2+\frac{4a^2n}{m_r}\int dk
\end{eqnarray}
Finally, the whole expression for the ground state energy can be
written in the integral form
\begin{eqnarray}
\label{energyfinal}
E^{ren}(\Omega,\omega)&=&\frac{3}{4}\frac{(\Omega-\omega)^2}{\Omega}
\\
&+&\alpha(1+z)\int_0^\infty
dl\left[1-\frac{1+z}{2}\frac{l^2}{\sqrt{1+4/l^2}}\int_0^\infty
ds\exp\left(-\frac{l^2\sigma_1(s)}{2}\right)\right]~,\nonumber
\end{eqnarray}
where
\begin{eqnarray}
\label{sigma11}
\sigma_1(s)=s\sqrt{1+4/l^2}+sz\frac{\omega^2}{\Omega^2}+\frac{z}{\Omega}\frac{\Omega^2-\omega^2}{\Omega^2}\left(1-e^{-\Omega
s}\right)~,
\end{eqnarray}
The integration in (\ref{energyfinal}) is performed over
dimensionless variables $s$ and $l$, and the energy
$E(\Omega,\omega)$ as well as the frequencies $\Omega$ and $\omega$
are now measured in the units $m_Bc^2$.

Following the principal of minimal sensitivity \cite{kleinertbook}
(or Feynman-Jensen variational inequality in case of the energy) we
seek the extremum point of the energy with respect to the
variational parameters $\Omega$ and $\omega$. Fig.~(\ref{zhelob})
shows the energy as a function of $\Omega,~\omega$ for the mass
ratio $z=4/3$, which corresponds ${\rm He}_3$ impurity in ${\rm
He}_4$ BEC, and for the value of coupling constant $\alpha=5$. This
dependence has the form of gutter having a minimum with respect to
$\Omega$ and not having it with respect to $\omega$.
\begin{figure}
\label{zhelob}
\includegraphics[width=17.0cm]{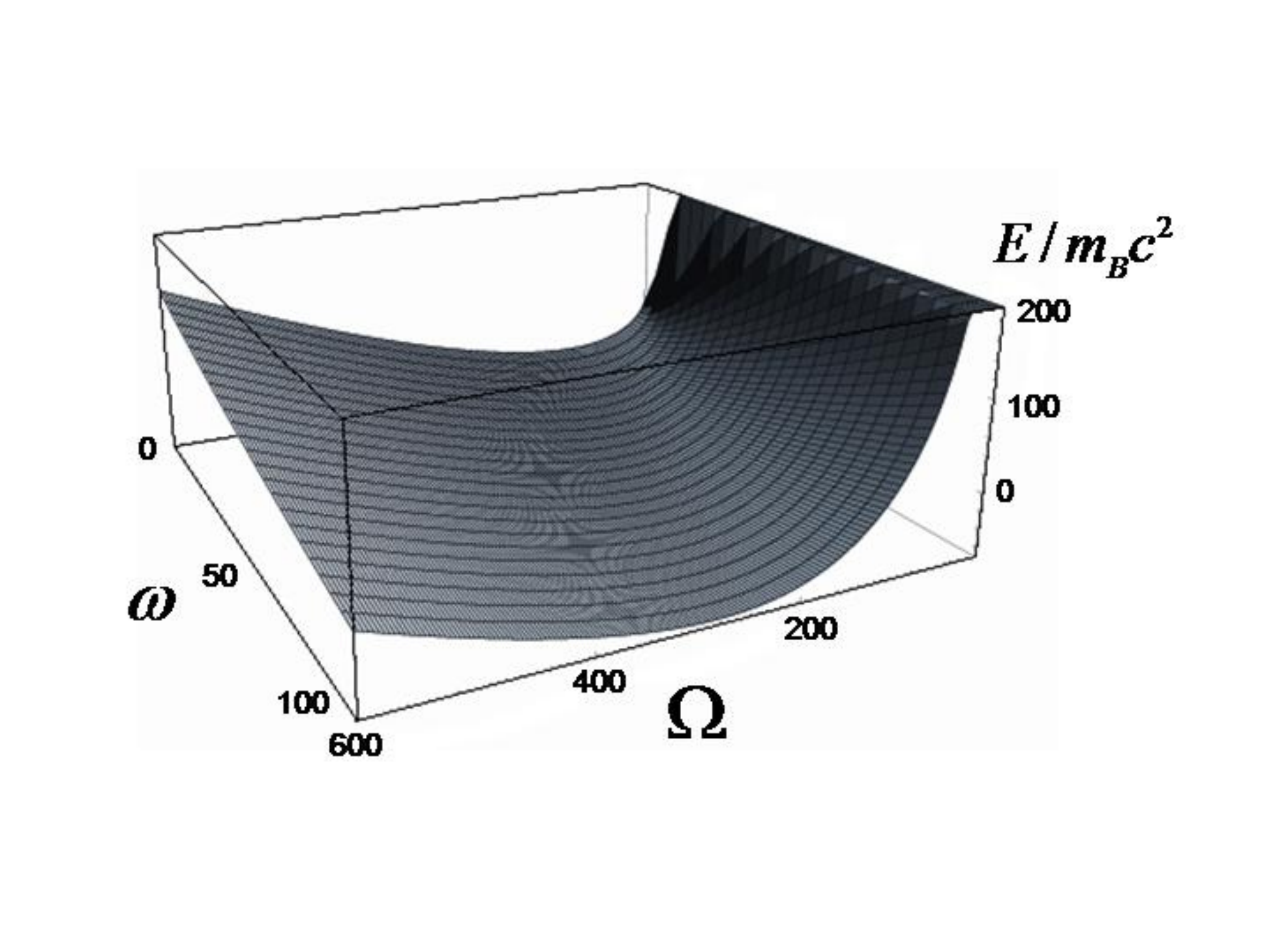}
\caption{ \label{zhelob} Energy as a function of variational
parameters $\Omega$ and $\omega$ as described by
Eq.~(\ref{energyfinal}). Alpha=1, z=4/3}
\end{figure}
As it can be seen from the Eq.~(\ref{energyfinal}), the strong
coupling asymptotic of the energy does not depend on $\omega$ and
behaves like $E_{strong}\sim-\alpha^2$ requiring the optimization
with respect to $\Omega$ only. On the other hand, the weak coupling
regime is independent of both variational parameters. Thus only the
intermediate coupling region is sensitive to the choice of the
second parameter $\omega$. On Fig.~(\ref{sample}) we plotted the
energy as a function of the coupling constant $\alpha$ optimized
with respect to $\Omega$ for different values of the second
variational parameter $\omega$. It turns out that at some value of
the coupling parameter $\alpha$ the energy becomes negative
indicating the existence of the bound impurity/BEC state in spite of
the repulsive interaction potential. One can see that the optimized
value of the energy rapidly converges to some minimum with
increasing $\omega$ becoming insensitive to $\omega$ at the values
of $\omega\geq 50$. Thus the use of this value is safe within
variational treatment. So the critical value of the coupling
constant $\alpha_c$ defined as $E(\alpha_c)=0$ decreases from
$\alpha_c\sim 3.2$ at $\omega\sim 1$ down to reliable value
$\alpha_c\sim 1.8$ at high $\omega$.
\begin{figure}
\label{sample}
\includegraphics[width=17.0cm]{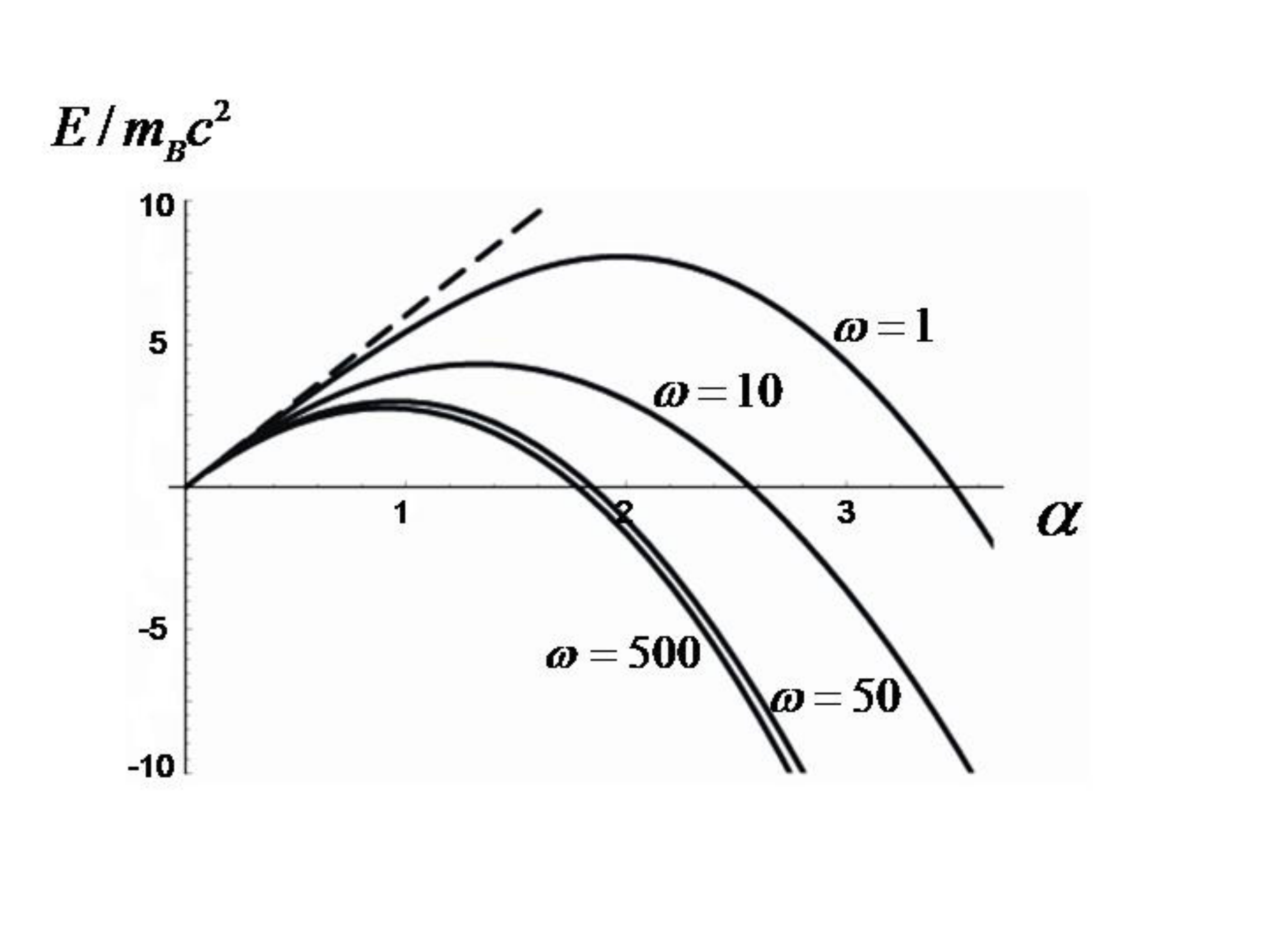}
\caption{ \label{sample} Energy optimized with respect to $\Omega$
as a function of the coupling constant $\alpha$ for different values
of second variational parameter $\omega$.}
\end{figure}
Next, Fig.~(\ref{critical}) shows the critical value of the coupling
constant $\alpha_c$ as a function of the mass ratio $m_B/M$. As one
would expect the impurity never self-localizes if its mass is much
bigger than the one of the Bose particle and binds with BEC in the
weak coupling regime in the opposite case of the small impurity
mass.
\begin{figure}
\label{critical}
\includegraphics[width=17.0cm]{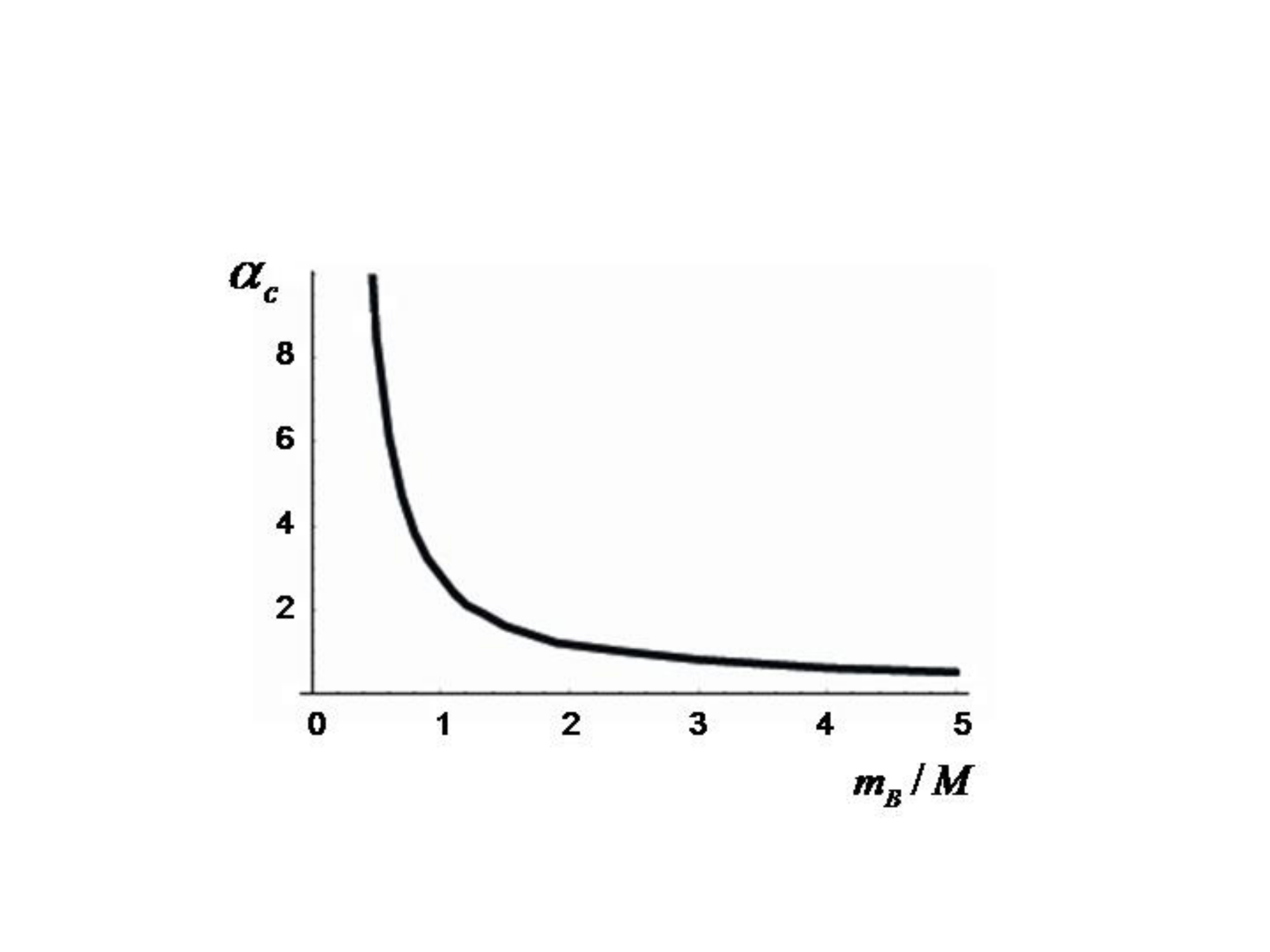}
\caption{  \label{critical} Critical value of coupling constant
$\alpha_c$ as a function of the mass ratio $m_B/M$.}
\end{figure}


\section{impurity correlation function}
Now let us directly consider the correlation function $C(x)$. Using
expression for the function $\Sigma_1$, Eq.~(\ref{sigma1}), one can
rewrite the expansion (\ref{ct}) up to first cumulant as
\begin{eqnarray}
C_T({\bf x},\beta\to\infty)=\exp\left[\beta
E_2-\frac{M\Omega}{8}\left(\frac{\Omega^2-\omega^2}{\Omega^2}\right)^2{\bf
x}^2+\frac{\Sigma_2({\bf x})}{W_0({\bf x})}\right]~.
\end{eqnarray}
The above function is normalized in such a way that $C_T({\bf
x}=0)=1$. Then, with the help of the
Eqs.~(\ref{generating},~\ref{exponenta}) the correlation function
can be written down in the integral form, namely
\begin{eqnarray}
\label{cor51} C_T({\bf
x},\beta\to\infty)&=&\exp\left\{-\frac{z\Omega}{8}\left(\frac{\Omega^2-\omega^2}{\Omega^2}\right)^2x^2+\alpha\frac{(1+z)^2}{2}\int_0^{\beta}
ds\int_0^sds^\prime\frac{l^2}{\sqrt{1+4/l^2}}\nonumber\right.
\\
&\times&\left.\exp\left(-\frac{l^2\sigma_1(s-s^\prime)}{2}\right)\left(\frac{\sin(lx\sigma_2(s-s^\prime))}{lx\sigma_2(s-s^\prime)}-1\right)\right\}~,
\end{eqnarray}
where the function $\sigma_1(s-s^\prime)$ is defined by the
Eq.~(\ref{sigma11}) and
\begin{eqnarray}
\sigma_2(s-s^\prime)&=&\frac{\omega^2}{\Omega^2}\frac{s-s^\prime}{\beta^\prime}+\frac{1}{2}\frac{\Omega^2-\omega^2}{\Omega^2}\frac{1}{\sinh\Omega\beta}
\\
&\times&\left(\sinh\Omega s-\sinh\Omega(\beta-s)-\sinh\Omega
s^\prime+\sinh\Omega(\beta-s^\prime)\right)~.\nonumber
\end{eqnarray}
As in Eq.~(\ref{energyfinal}), the integration on the right hand
side of the above expression is performed over dimensionless
variables $s,~s^\prime$, and $l$, temperature $1/\beta$ and
freqiencies $\Omega$ and $\omega$ are measured in units $m_Bc^2$
while the variable $x$ is defined in units $1/m_Bc$. Finally, in
order to find the best approximation for the correlation function
the expansion (\ref{cor51}) has to be optimized with respect to the
pair of the variational parameters $\Omega,~\omega$. In general, if
the optimization procedure is directly applied to the expansion for
the correlation function, the optimized values of variational
parameters should depend on the coordinate ${\bf x}$. However, it
turns out that instead of the variational optimization of the
coordinate-dependent correlation function one can use the optimal
values of the variational parameters obtained from the optimization
of energy. These coordinate-independent values then has to be
substituted into the expansion for the correlation function giving
the variationally improved result, i.e. convergent strong coupling
expansion for the correlation function. This scheme of the
variational perturbation theory was successfully applied for the
variational solutions of time-dependent equations in non-linear
dynamics \cite{pelster2006, novikovpelster}.

We will also define the mean value of $x^2$ representing the square
of the radius of localization or polaron radius as
\begin{eqnarray}
\langle x^2\rangle=\frac{\int x^2C({\bf x})d{\bf x}}{\int C({\bf
x})d{\bf x}}
\end{eqnarray}

\begin{figure}
\label{correlator}
\includegraphics[width=17.0cm]{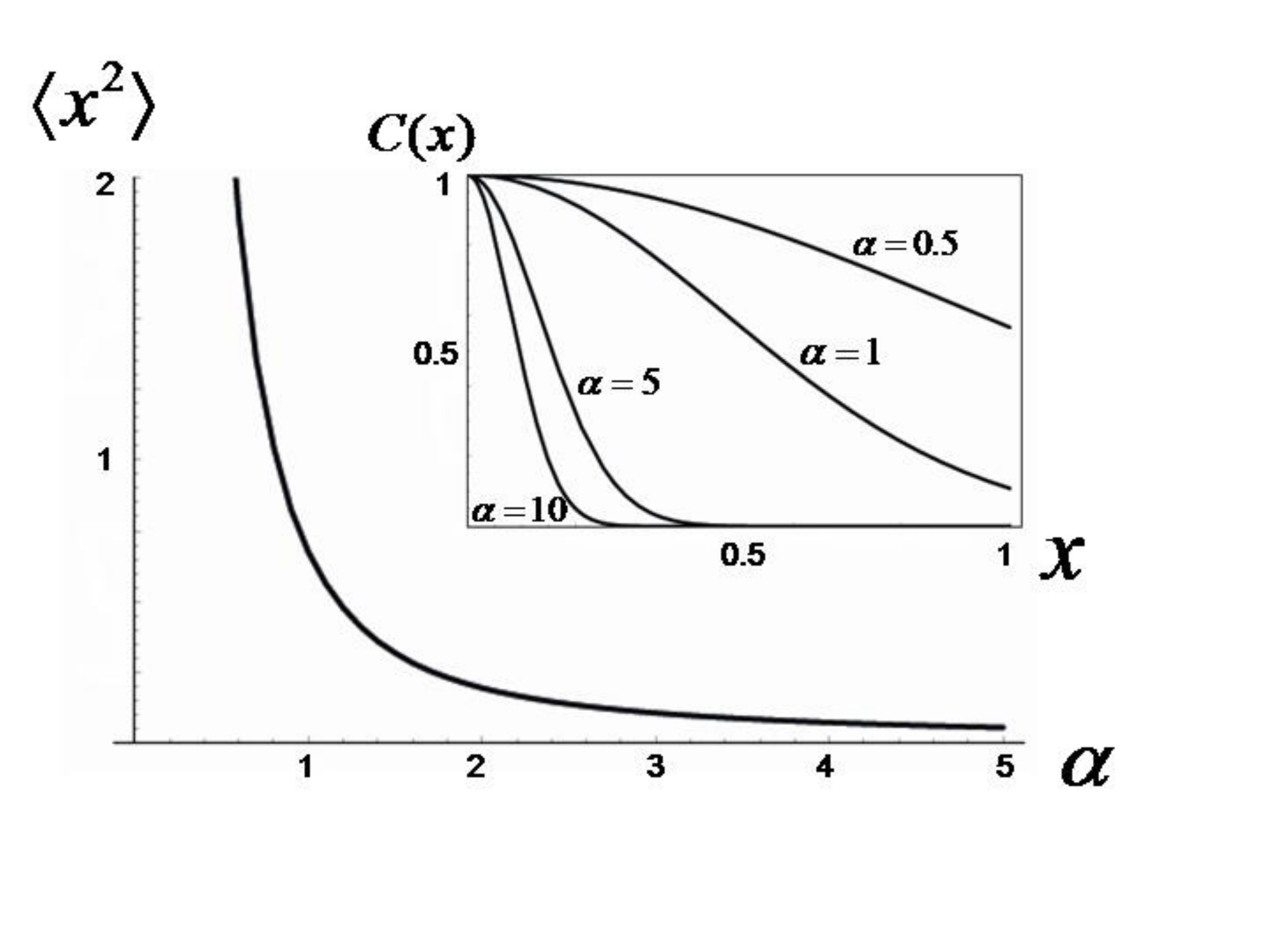}
\caption{ \label{correlator} Mean value of $x^2$ as a function of
coupling constant, x is shown in units $1/m_Bc$. Inset - correlation
function $c(x)$ for different values of coupling constant
$\alpha=0.5$ (upper curve), $\alpha=1,5$ and $\alpha=10$ (lower
curve). }
\end{figure}
Fig.~(\ref{correlator}) shows the mean square of the radius of
localization as a function of coupling constant. It has inverse
dependence on coupling strength.  The inset shows the correlation
function itself for different values of coupling strength.  The
correlation function has Gaussian-like shape regardless of $\alpha$
which means that formally the particle is always localized.  This is
the consequence of the choice of the trial action used in this
method.  However, this correlation function should converge to the
true-one at large values of interaction strength. It is interesting
to examine the localization radius at the coupling constant
$\alpha_e$ at which the energy has maximum, i.e. $dE(\alpha)/d\alpha
|_{\alpha_e} = 0$.  We notice that
$\langle x^2(\alpha_e) \rangle \sim 1/m_B c~$.
This can be expressed through a so called healing length of the BEC
$\xi =1/zp_c$, where $p_c$ is the critical momentum of the impurity
above which the dissipation takes place in case of real-time
dynamics in accordance to Landau's criterion.  This leads to
$\sqrt{\langle x^2(\alpha_e) \rangle} \sim 1/p_c$, or using
Heisenberg uncertainty for the localized particle $\sqrt{\langle p^2
\rangle} \sim p_c$, i.e.~the average momentum of a particle at the
point of self localization is of the order of the critical momentum.
One can make physical sense of this fact by noticing that this means
that the tendency of self-localization appears when the exchange of
energy between particle and BEC becomes possible.

\section{CONCLUSION}
\label{conclusion}

In this work we consider a problem of self-localization of impurity
in degenerate BEC.  This problem has been considered previously
using the mean-field approach.  Here we use a full quantum
description of the ground state of impurity surrounded by BEC. The
variational perturbation method is employed to calculate the
imaginary-time propagator of impurity in degenerate BEC.  The free
energy and the spatial correlation of impurity in BEC is obtained as
a function of coupling strength.  Our results point to a possible
relation between self localization and the real-time particle/BEC
energy exchange. In this work we explicitly used the degeneracy of
the BEC. However, the methodology developed in this paper can be
extended to Bose systems without an assumption of diluteness.  The
latter can be done by using the four-point Green's function of the
strong coupled Bose liquid \cite{bogoliubov2003} in the leading
expansion term instead of degenerate BEC propagator used in this
work.

\section{acknowledgments}
This work has been supported by the NSF CAREER award ID 0645340.

\end{document}